\documentclass[useAMS,usenatbib,usegraphicx]{mn2e}
\usepackage{amsmath,amsfonts,amssymb}

\newcommand{\lya}{Ly$\alpha$}

\title[Galactic Winds]{Semi--Analytic Simulations of Galactic Winds: Volume
Filling Factor, Ejection of Metals and Parameter Study}

\author[Bertone, Stoehr \& White]
{Serena Bertone$^{1,2}$\thanks{E-mail: serena@mpa-garching.mpg.de},
Felix Stoehr$^{1,3}$ and Simon D.M. White$^{1}$ \\
$^{1}$Max Planck Institut f\"ur Astrophysik, Karl Schwarzschild Str. 1,
85741 Garching bei M\"unchen, Germany  \\
$^{2}$Dipartimento di Fisica Generale, via P. Giuria 1, 10125 Torino, Italia \\
$^{3}$Institute d'Astrophysique de Paris, 98bis, bd Arago, 75014 Paris, France}

\begin{document}

\date{Submitted to MNRAS}

\pagerange{\pageref{firstpage}--\pageref{lastpage}} \pubyear{2003}

\maketitle

\label{firstpage}

\begin{abstract}
We present a semi--analytic treatment of galactic winds within high resolution,
large scale cosmological N--body simulations of a $\Lambda$CDM Universe.
The evolution of winds is investigated by following the expansion of supernova
driven superbubbles around the several hundred thousand galaxies that form
in an approximately spherical region of space with diameter $52 h^{-1}$ Mpc
and mean density close to the mean density of the Universe.
We focus our attention on the impact of winds on the diffuse intergalactic
medium. Initial conditions for mass loss at the base of
winds are taken from \citet{shu}.
Results are presented for the volume filling factor and the mass fraction of the
IGM affected by winds and their dependence on the model parameters is carefully
investigated.
The mass loading efficiency of bubbles is a key factor to determine the
evolution of winds and their global impact on the IGM: the higher the mass
loading, the later the IGM is enriched with metals.
Galaxies with $10^{9} < M_{\star} < 10^{10}$ M$_{\sun}$ are responsible for most
of the metals ejected into the IGM at $z=3$, while galaxies with $M_{\star} <
10^{9}$ M$_{\sun}$ give a non negligible contribution only at higher redshifts,
when larger galaxies have not yet assembled.
We find a higher mean IGM metallicity than \lya\ forest observations suggest
and we argue that the discrepancy may be explained by the high temperatures of
a large fraction of the metals in winds, which may not leave detectable imprints
in absorption in the Ly$\alpha$ forest.

\end{abstract}


\section{Introduction}

Powerful outflows from star--forming galaxies have been detected
throughout the history of the universe (\citealt{heckman90}, \citealt{hlsa},
\citealt{adelberger}), providing, perhaps, the mechanism to transport metals
from the interstellar medium (ISM) of galaxies to the low density intergalactic
medium (IGM). This could at least partially explain the widespread level of
chemical enrichment observed in the spectra of quasars (\citealt{cowie},
\citealt{joop}, \citealt{ell}, \citealt{simcoe}).

The energy necessary to power outflows on galactic scales is supplied by
supernova explosions and winds from young massive stars in OB associations.
Any episode
of star formation may create a superbubble in the ISM and, if the rate of
energy input is large enough, the superbubble can blow out of the ISM and
create a wind. In local starbursts (\citealt{phil}, \citealt{cecil2},
\citealt{walter}, \citealt{sugai}), winds have been observed to extend to at
least 10 kpc from their host galaxies. \citet{ses} claim that winds can reach
even larger distances, but are unobservable because of the low emissivity of the
outflowing gas.

At present, it is difficult to predict which galaxies are responsible for
seeding the IGM with metals or to establish the effects that supernova--driven
blastwaves have on the galaxy formation process.
While gravity does not influence the evolution of superbubbles in the ISM, it
is crucial for determining the long term fate of winds. Since winds from dwarf
galaxies form in shallower potential wells, they are the most likely to
be able to disperse their metal content into the IGM. On the other hand,
\citet{ses} suggest
that most of the energy from winds resides in a hot ($T\sim 10^7$
K) low density component that can escape the galaxies even when the bulk of the
outflowing mass is retained. \citet{mac}
demonstrate that metals are easily accelerated to velocities larger than the
escape velocity, implying that a galaxy can lose a high fraction of
its metals even with a relatively low mass ejection efficiency.
Although winds may occur more frequently in dwarf galaxies, the metals ejected
by massive galaxies may dominate the total budget.
It is therefore
not a trivial problem to assess which galaxies have been responsible for the
pollution of the IGM and when the enrichment occurred.

Several groups have applied simple phenomenological prescriptions to simulations
in order to investigate the effects of winds on the IGM and some important
results have emerged. For example, \citet{mfr} find that pregalactic
outflows are an efficient mechanism for distributing the metals produced in
stars over large cosmological volumes, prior to the reionisation epoch.
\citet{aguirre} argue that radiation pressure ejection or winds from relatively
large galaxies at lower redshifts can account for the observed metallicity of
the IGM and the intracluster medium. 
\citet{tom}  and \citet{croft} demonstrate that cavities evacuated by winds in
the outskirts of galaxies may leave characteristic signatures in the
Ly$\alpha$ forest.
In contrast, \citet{tv} find that winds have little effect on the statistics of
H I absorption lines and produce C IV absorption lines in reasonable agreement
with observations.

The significance of galactic winds for the evolution of the IGM is still not
fully established, however.
Both hydrodynamic and semi--analytic simulations use
phenomenological prescriptions for the physics of galactic winds and new
parameters have to be introduced to account for the
uncertainties that derive from a still uncomplete observational picture.
In particular, no well founded relation is available to link the properties of
the ISM and the morphology of its host galaxy to the structure and evolution of
the outflows.
Because of insufficient resolution and incomplete physics, numerical results
often disagree with each other and the effects of winds on
the Ly$\alpha$ forest remain controversial, leaving the way open for further
studies.

In this paper, we present a new implementation of the physics of galactic winds
within the semi--analytic galaxy formation model of \citet{semi}, and we apply
it to a set of high resolution N--body simulations of structure formation in a
$\Lambda$CDM universe (\citealt{felix}, \citealt{bene}).
By using a high resolution simulation of a spherical region of diameter
52 $h^{-1}$ Mpc, we investigate the long term evolution of winds and their
effects on a typical region of the IGM.
We solve the equation of motion for a spherical astrophysical blastwave
to follow the evolution of winds after they escape the visible regions of
galaxies.
Our phenomenological model for winds uses the initial conditions
proposed by \citet{shu}, which parameterise the mass loss and the initial
velocity of winds as a function of the star formation rate of the galaxy.
Here we follow the evolution of galactic winds throughout most of the
history of the universe and we outline their impact on the IGM by estimating the
fraction of volume and mass of the IGM which they affect as a function of
time and model parameters.

This paper is organised as follows: in Section \ref{due} we present our set of
high resolution N-body simulations and the semi--analytic prescriptions we
adopt to model the physics of galactic winds; in Section \ref{tre} we
outline some global properties of winds as a function of the model parameters;
in Sections \ref{quattro} and \ref{budget} we present the results for the volume
filling factor and the fraction of mass affected by winds and
in Section \ref{ejection} our findings for the ejection of metals into
the IGM;
finally, in Section \ref{mares} we discuss the dependence of our results on the
numerical resolution of the simulations and we draw our
conclusions in Section \ref{cinque}.

\section{Simulating Galaxy Formation and Feedback Processes}
\label{due}

\subsection{The N--body Simulations}

Here we describe our set of high resolution N--body
simulations. Later subsections present the
prescriptions we adopt for the physics of galactic winds and their numerical
implementation.
We assume a $\Lambda$CDM cosmology with matter density
$\Omega_m =0.3$, dark energy density $\Omega_{\Lambda} =0.7$, Hubble constant
$h=0.7$, primordial spectral index $n=1$ and normalisation $\sigma_8 = 0.9$.

The use of pure N--body simulations allows us to find
a good compromise between high mass resolution and a large simulated volume,
although this choice implies that the physics of baryons cannot be
followed directly.
A high resolution in mass is crucial to determine the role of galaxies with
different masses in polluting the IGM with metals, while a large region is
necessary to study the effects of winds in their proper cosmological context.

Our simulations are resimulations at higher resolution of a ``typical''
spherical region with a diameter of approximately 52 $h^{-1}$ Mpc and average
density close to the cosmic mean. About half of the enclosed galaxies at $z=0$
are field galaxies, while the rest are in groups and poor clusters.
The simulated region was identified within the much larger cosmological ``VLS''
simulation run by the VIRGO Consortium (\citealt{jenkins}, \citealt{b1}).
It was resimulated four times with increasing internal mass resolution and
decreased external resolution. The effects of the large
scale gravity field on the region of interest are thus correctly retained.
The mass of the dark matter particles in the high resolution region of ``M3'' is
$1.7\cdot 10^8 h^{-1}$ M$_{\sun}$ and the number of particles about $7\cdot
10^7$.
The initial conditions were generated with  ZIC \citep{zic} and the 
simulations were performed using the parallel treecode {\small GADGET}
\citep{gadget}. The dark matter evolution was followed from redshift $z=120$
down to redshift $z=0$ and 52 simulation outputs were stored between $z=20$ and
$z=0$.

\subsection{Galaxy Formation and Star Formation History}
\label{sfhistory}

The formation and evolution of galaxies is modelled with the semi--analytic
technique proposed by \citet{kauffmann} in the new implementation by
\citet{semi}. Merging trees extracted from the simulations are used to
follow the galaxy population in time, while simple prescriptions for gas
cooling, star formation and galaxy merging model the processes
involving the baryonic component of the galaxies. Both the spectrophotometric
evolution of the stellar population and the morphological evolution of the
galaxies can thus be modelled in detail.

Dark matter haloes and subhaloes are identified with the algorithm
{\small SUBFIND} \citep{semi} and a catalogue is compiled with all the groups
and subhaloes that contain at least ten particles, meaning that for M3 the
minimum dark matter mass of a subhalo is $1.7\cdot 10^9 h^{-1}$ M$_{\sun}$.
At $z=3$ a total of about four hundred thousand galaxies are
identified and about three hundred fifty thousand are present at $z=0$.
The two largest clusters, each with a total mass of about $10^{14} h^{-1}$
M$_{\sun}$, assemble most of their mass after $z\sim 1$.

The convergence of the star formation history in the ``M'' series of simulations
has been investigated by \citet{bene}. They show
that the lower the mass resolution, the later in time the simulations are able
to account for all the star formation in the region, since at redshifts higher
than $z\sim 10$ the major contribution comes from objects with total masses of
order $10^9$ M$_{\sun}$, while only at lower redshifts do more massive objects
appear and become dominant. By
comparing the results of the star formation history of M3 with a higher
resolution cluster simulation by \citet{semi}, Ciardi and collaborators estimate
that M3 is able to account for most of the star formation at $z\la 11$.

\subsection{Feedback Prescriptions}

In our model, new recipes for mechanical feedback from supernovae are
introduced in order to include the physics of galactic winds.

\cite{kauffmann} and \cite{semi} find that a simple recipe for feedback,
implemented in the so--called ``ejection'' scheme, is sufficient to give
reasonable predictions for some observed properties of galaxies, e.g. the
suppression of star formation in low mass haloes and the slope of the
Tully--Fisher relation. However, the simplicity of this prescription makes it
impossible to follow in detail the evolution of galactic winds.
In particular, the scheme does not describe the diffusion of the matter and
metals lost by galaxies, because there are no recipes for following the
evolution of wind ejecta. This is what we aim to provide in this paper.

Here we use the semi--analytic model of \citet{semi} with the implementation of
the ejection scheme, to follow the evolution of the cold gas and the stellar
component of the galaxies.
We add our recipes for winds on top of this pre--existing scheme, without
modifying its prescriptions.
This is not fully consistent, because we do not modify the cooling and the
infall prescriptions in the semi--analytic code to match our new model for the
immediate surroundings of galaxies.
One consequence of this is that the total metal and gas mass in our
simulated region is not exactly conserved. However, violations are minor.

Since we want to investigate the effects of winds on the IGM by applying our
model to a large simulated region, we are neither able nor interested to
resolve the details of the first phases of wind evolution, when the
superbubbles blow out of the ISM of galaxies.
Nor do we model the impact of the outflow on the physical
conditions of the ISM in the host galaxy. Here, we are concerned with
the long--term evolution of the winds once they have escaped the visible regions
of galaxies.

We make the simplifying assumption of spherical symmetry for the wind evolution.
Galactic outflows observed in nearby galaxies appear to be mostly bipolar
\citep{hlsa}, with the gas escaping preferentially along the direction where the
gravitational potential gradient is steeper.
However, observations of high redshift objects (\citealt{fbb},
\citealt{pettini}) suggest that most galaxies are affected by large scale winds
implying near spherical outflows. Together with the fact that an initially
nonspherical bubble approaches sphericity at later times \citep{omk}, this
suggests that our symmetry assumption may be appropriate.

The thermal energy injected by supernova explosions is converted
to kinetic energy and the outflow remains approximately adiabatic until
radiative losses become substantial.
During the adiabatic phase of the evolution,
radiative losses are negligible and the expansion of the bubble is driven by
the pressure of the hot plasma, which acts as a piston on the surrounding
medium.
The outflow can be described as an adiabatic blastwave expanding into a
cosmologically structured context, and its dynamics obeys a virial
theorem as stated by \citet{omk}.
This phase typically happens during the early evolution of winds immediately
after blow out, when a newly formed bubble starts expanding into the galactic
halo,
the mass of ejecta is larger or comparable to the mass of the swept up gas and
the cooling time of the hot gas within the bubble is longer than the
age of the wind.
Observationally, \citet{hoopes} and \citet{ses} have proved that the energy lost
through radiative cooling of the coronal ($T \sim 10^{5.5}$ K) and the hot ($T
\sim 10^7$ K) phases of the wind in the starburst galaxy M82 is small,
supporting the idea that the early evolution of this wind is nearly adiabatic.

The interior of a hot bubble is made up of two phases: a hot phase which we
assume to have roughly uniform temperature and pressure, and 
dense gas falling in from surrounding ``filaments''.
In our model, the hot phase is a mixture of shocked wind gas, of shocked
low density ambient gas and of denser ambient gas stripped from the
infalling gas clouds. The shocked
wind gas is in turn made up of SN ejecta and of entrained interstellar medium
outflowing from the galaxy.
In their SPH simulations, \citet{sh} find that winds expand
anisotropically into low density regions and that although a significant
fraction of the gas from infalling filaments is ``entrained'' (i.e. mixed into the hot
phase) most remains cool and dense.
\citet{sh} also find no clear radial stratification of the phases
within the bubble: the supersonic outflow region of the wind fills a very
small volume near the galaxy and there is no clear separation between
shocked wind gas, shocked diffuse ambient gas, and initially denser
ambient gas ``entrained'' into the hot phase after filaments are
engulfed by the bubble. No significant cool, dense shell forms near the outer
shock radius.

The adiabatic phase is terminated when the loss of energy by radiation
becomes substantial, that is most of the energy transferred to the swept up gas
is radiated away and the total energy content of the hot bubble decreases. This
phase sets in when the cooling time of the hot bubble becomes shorter than the
age of the wind. At this point, a thin shell of cooled gas forms near the
bubble's outer boundary and continues to expand pushed by the momentum input
from the wind.

In our simulations, we model galactic winds as uniform pressure--driven bubbles
of hot gas emerging from star forming galaxies, which evolve adiabatically until
their cooling time becomes shorter than their dynamical expansion time.
After this moment, we switch to a momentum--driven approximation and
we assume that the mass swept up by the wind accumulates in a thin cooled
shell pushed by the momentum accreted from the wind.
During this second phase, both the thin shell and the bubble interior are cool.

\subsubsection{The Adiabatic Phase: Pressure--Driven Bubbles}
\label{adiabatic}

According to \citet{omk}, under the assumption of negligible energy losses, the
equation for the conservation of energy of a spherical bubble with energy
injection at the origin is
\begin{eqnarray}\label{energy}
   \frac{d E}{dt}&=& \frac{1}{2}\dot{M}_w v_w ^2 +
   \varepsilon 4\pi R^2 \cdot \nonumber \\
   & & \left\{ \left[ \frac{1}{2}\rho_o v_o ^2 +
   u_o - \rho_o \frac{GM_h}{R}
   \right] \left( v_s - v_o \right) - v_o P_o \right\} .
\end{eqnarray}

Here $R$ and $v_s$ are the radius and the velocity of
the shock, $\dot{M}_w$ and $v_w$ the mass outflow rate and the outflow velocity
of the wind, $\rho_o$, $P_o$ and $v_o$ the density, the pressure and the outward
velocity of the surrounding medium, $M_h$ the total mass internal to the shock
radius and $\varepsilon$ a parameter, the entrainment fraction, defining the
fraction of mass that the bubble sweeps up while crossing the ambient medium.
The first term on the right hand side is the energy injected by the
starburst, while the terms in brackets represent the energy variation due to the
accretion of gas onto the bubble. The newly entrained gas mass contributes
its kinetic ($\propto \rho_o v_o ^2$), internal ($\propto u_o$) and potential 
($\propto G M_h$) energy to the total energy in the bubble, but it requires work
to be accelerated against the pressure forces of the surrounding medium
($\propto P_o$).
We neglect the gravitational energy transfer to the dark matter component, which
is small and does not significantly change the energy budget.
For the conservation of mass law, the total mass in the bubble $M_b$ is the sum
of the outflowing wind mass plus the swept--up mass:
\begin{equation}\label{monster1}
\frac{dM_b}{dt}=\dot{M}_w + \varepsilon 4\pi R^2 \rho_o \left( v_s - v_o\right).
\end{equation}
The radius of the shocked bubble is given by $v_s=dR/dt$.

At blow out, that is when the wind escapes the galactic disk or spheroid, we
assume that a bubble is formed initially with a radius equal to the galaxy
radius $R_g$ ($R_o=R_g$), no mass ($m(R_o)=0$) and velocity equal to the wind
velocity ($v_s(R_o)= v_w$).
After blow out, the bubble starts to accumulate gas. Since our semi--analytic
model does not follow the internal structure of galaxies, we have to make a
further assumption for the galaxy radius, in order to link it to the properties
of the dark matter halo in which the galaxy is embedded.
Thus we fix $R_g$ to be a given fraction of the
virial radius of the DM halo, e.g. $R_g = r_{200}/10$. This choice gives values
in rough agreement with the observed radii of galaxies at all redshifts.

Equations for the evolution of winds in the thin shell approximation have
previously been used in similar work by \citet{tom} and \citet{aguirre}.
As noted by \citet{omk}, the approximation of a thin shell holds for radiatively
cooled blastwaves and for blastwaves expanding in the Hubble flow, but it is
not applicable to adiabatic blastwaves evolving in a static or infalling medium.
Beside the different formalism used to describe the winds, we make different
assumptions for the initialisation of the bubble properties, which we think are
more realistic than those of \citet{tom} and \citet{aguirre}.
\citet{tom} fix the initial conditions at the virial radius and
assume that the initial shell mass is equal
to $m = \left( \Omega_b / \Omega_m \right) M_{vir}$. This choice implies that
winds blow out all the baryonic mass of the galaxy, including its stars.
The evolution of winds in the outskirts of galaxies immediately
after blow out is crucial for determining their ability to escape the
gravitational attraction of haloes.
Thus, following the evolution of winds only for $R > R_{vir}$ may neglect an
important stage in their formation which could significantly affect the
reliability of the final results.
\citet{aguirre} set initial conditions for the shell mass by choosing a radius
$R_o$ to include a fixed fraction $\xi$ of the galaxy mass.
The initial shell mass is therefore $m(R_o ) \propto \xi M_{gal}$, but it is not
clear why, when computing wind evolution, a significant fraction of the
galaxy mass should be assumed to be already in the shell when the wind emerges from the
galaxy.

During the phase of adiabatic expansion, the pressure of the hot shocked bubble
can be expressed as a function of the bubble energy, that is
\begin{equation}
P_b = \frac{E}{2\pi R^3}.
\end{equation}
Given the bubble pressure $P_b$, a simple estimate of the bubble temperature is
then
\begin{equation}\label{temp}
T_b = \frac{\mu m_H E}{2 \pi k \rho_b R^3},
\end{equation}
with $\mu$ the mean molecular weight, $m_H$ the mass of atomic hydrogen and
$k$ the Boltzmann constant.
At blow out, most of the bubbles have temperatures in excess of
$10^6$ K. During the subsequent adiabatic evolution, the bubble temperature
is determined by two competing processes: it decreases because of the adiabatic
expansion of the bubble and it increases because of the energy injected by the
starburst.

The adiabatic expansion continues until the cooling time of the hot shocked
bubble becomes shorter than the age of the wind.
We calculate the cooling time of bubbles as a function of their temperature
$T_b$, mean density $\rho_b$ and metallicity $Z_b$ as:
\begin{equation}\label{tcool}
t_{cool} = \frac{3\rho_{b}kT_{b}}{2 \mu m_H n^2 \Lambda\left( T_b, Z_b \right)},
\end{equation}
where $n$ is the total ion density and $\Lambda \left( T_b, Z_b \right)$ is the
metal dependent cooling function of \citet{sd}.
For most bubbles at blow out, the cooling time is at least one order of
magnitude larger than the age. When the wind expands into a high density
environment, the cooling time decreases steadily with time and soon the wind
makes the transition to a momentum driven shell.
This happens particularly often at
high redshift, where the mean density of the Universe is higher and the energy
provided by star formation smaller. Another important factor that determines the
shortening of the cooling time is the mass loading of winds: the higher the mass
loading efficiency, the higher the bubble density, the shorter the cooling time.
We find that the cooling time of a bubble normally becomes shorter than the
dynamical expansion time after the swept-up mass exceeds the mass of wind
ejecta.

\subsubsection{The Radiative Phase: Momentum--Driven Shells}
\label{radiative}

When a bubble becomes radiative, the expansion work done on the ambient medium
is radiated away, at the expenses of the total energy of the bubble.
From this point onward, the dynamics of the wind is dominated by the momentum
imparted onto the thin shell of cooled material by the outflowing hot gas.
The equation of motion for the spherically symmetric thin shell
that accumulates mass at the shock radius is given by the conservation of
momentum :
\begin{eqnarray}\label{label}
   \frac{d}{dt}\left( m v_s \right) & = & \dot{M}_w \left( v_w - v_s \right) -
   \frac{GM_h}{R^2} m - \nonumber \\
   & & - \varepsilon 4\pi R^2 \left[ P_o + \rho_o v_o
   \left( v_s - v_o\right) \right],
\end{eqnarray}
with $m$, $R$ and $v_s$ the mass, the radius and the velocity of the shell
respectively. The first term on the right--hand 
side of the equation represents the momentum injected by the starburst,
the second term takes into account the gravitational attraction of the dark
matter halo and the two final terms represent the thermal and the ram pressure
of the surrounding medium.
The conservation of mass law gives
\begin{equation}\label{monster2}
\frac{dm}{dt}=\dot{M}_w \left( 1- \frac{v_s}{v_w} \right) +
              \varepsilon 4\pi R^2 \rho_o \left( v_s - v_o\right),
\end{equation}
while the radius of the shell is again given by the equation $v_s=dR/dt$.

\subsubsection{Wind Velocity and Mass Loss Rate}

At present, both semi--analytic and SPH simulations use empirical prescriptions
for the physics of galactic winds and the velocity and the mass outflow rate
are assumed as parameters (e.g. \citealt{sh}, \citealt{aguirre},
\citealt{tv}, \citealt{violence}).
This approach has proved useful, although the simulated results
depend sensitively and in a complex fashion on the choice of the parameters.

\citet{shu} proposed a more detailed model that links the wind to the
star formation properties of galaxies.
They start from two observational facts: (i)
the outflow rate in galaxies at every redshift is of the order of the star
formation rate \citep{martin} and (ii) the initial wind velocities seem to be
independent of the galaxy morphologies (\citealt{hlsa}, \citealt{fbb})
and lie in the range 100--1500 km s$^{-1}$.
By using the theoretical models of \citet{mko} and \citet{ef}, they predict the
mass outflow rate $\dot{M}_w$ and the wind velocity $v_w$ at blow out as
a function of the star formation rate $\dot{M}_{\star}$ of the host galaxy,
\begin{equation}\label{cond1}
\dot{M}_w = 133 \left(
\frac{\dot{M}_{\star}}{100\textrm{M}_{\sun}\textrm{ yr}^{-1}} \right)^{0.71} K 
\textrm{ M}_{\sun}\textrm{ yr}^{-1},
\end{equation}

\begin{equation}\label{cond2}
v_w = 623  \left(
\frac{\dot{M}_{\star}}{100\textrm{M}_{\sun}\textrm{ yr}^{-1}} \right)^{0.145}
K^{-1/2} \textrm{ km s}^{-1},
\end{equation}
where $K$ is a constant that takes into account various properties of the ISM.
It depends on the efficiency of conduction relative to the thermal
conductivity of clouds, on the minimum radius of clouds in the ISM and on the
dimensions of star--forming regions (see \citealt{shu} for a comprehensive
discussion).
In the following, we will call the ratio between the
wind mass loss rate and the star formation rate the ``ejection rate''
$\left( R_{e} = \dot{M}_w / \dot{M}_{\star} \right)$ of the wind.

Note that the momentum input $\dot{M}_w v_w$ in this model is only weakly
dependent on $K$ ($\propto K^{1/2}$), with a stronger dependence on the
star formation rate ($\propto \dot{M}_{\star}^{0.855}$). The energy
input per unit of star formation rate is completely independent
of $K$ and therefore of all other galaxy properties.
\citet{shu} give a number of arguments in support of this very simple model
which is quite similar to the earlier model of \citet{dekel}.

The theoretical predictions can be fine--tuned to reproduce the observations
with reasonable accuracy both for the mass loss rate and the wind velocity.
In order to make our predictions consistent with the observations of
\citet{martin}, we fix a maximum value for the ejection rate of
$R_{e}^{max} = 5$. Equations (\ref{cond1}) and (\ref{cond2})
tend to overestimate $R_{e}$ for low values of the star formation rate.
In the following, we choose as our fiducial value $K= 0.5$ and we investigate
two more models with $K=0.1$ and $K=1$. Since the overall effect of $K$ on the
results is mostly weaker than the one of the entrainment fraction, we
concentrate our analysis on the other parameter. However, we remind the reader
that variations in $K$ do produce an appreciable degeneracy in the results.

\subsubsection{Metals in Winds}

The metallicity of the wind fluid depends both on the
amount of metals ejected by supernovae and on the metallicity of the ISM.
In fact, the mass ejected by winds is the sum of two components: the metal
enriched stellar ejecta from supernova explosions and the shocked
ISM entrained in the outflow. The latter represents the major
fraction of the mass lost by the galaxy, constituting about
90\% of the ejecta, for a mass loss rate comparable to
the star formation rate.
Assuming for star formation a mass yield $Y=0.2$, corresponding to the fraction
of mass converted into stars that is returned to the ISM by supernova
explosions, then the mass of outflowing gas which is
entrained ISM is $\dot{M}_{\star} dt \left( R_{e} -Y\right)$.
Similarly, the metal mass in the wind fluid is the sum of the metals ejected by
supernovae, whose metal yield is $Y_z$, and the metals in the shocked ISM
\begin{equation}
\dot{M}_{z,w}=\dot{M}_{\star} \left[ Y_z +\left( R_{e} -Y\right) Z_{ISM}\right],
\end{equation}
where $Z_{ISM}$ is the metallicity of the ISM.
In our semi--analytic model, galaxies are schematically represented as a disc of
cold gas, which constitutes the ISM of the galaxy, surrounded by a halo of hot
gas.
The total mass of metals accreted by a bubble during its adiabatic expansion
reflects the form of the mass conservation equation (\ref{monster1}) and is the
sum of the metals accreted from the wind fluid and the metals accreted from the
ambient medium:
\begin{equation}
\dot{m}_z = \dot{M}_{z,w} + \varepsilon 4\pi R^2 \rho_o
   \left( v_s - v_o \right) Z_{hot},
\end{equation}
with $Z_{hot}$ the metallicity of the hot gas.
The second term indicates the amount of metals swept up by the wind in the
halo of the galaxy or of the galaxy group and does not give any contribution for
bubbles that are expanding far into the IGM, since the IGM itself is assumed to
contain no metals.

The metal mass accreted by a shell during the momentum driven expansion reflects
the form of equation (\ref{monster2}) and contains one more multiplicative
factor that takes into account the actual amount of wind material that is
accreted onto the shell:
\begin{equation}
\dot{m}_z = \dot{M}_{z,w} \left( 1 - \frac{v_s}{v_w} \right) + 
            \varepsilon 4\pi R^2 \rho_o \left( v_s - v_o \right) Z_{hot}.
\end{equation}

\subsubsection{The Wind Environment}
\label{environment}

Once a wind is formed, it expands through the halo of its host galaxy and, if
it is energetic enough, it can escape the gravitational attraction of the halo
and break out into the IGM.
The winds attached to galaxies in groups are subject to the
gravitational field of the group and the closer a galaxy lies to the centre of
a massive group, the more energetic the wind has to be to be able to escape
the potential well.

When simulating the evolution of the winds, it is therefore important to
know the density distribution of the gas into which the winds expand. Our
semi--analytic prescriptions provide this information by assuming that inside
dark matter haloes the gas follows the distribution of the dark matter.
We model the gravitational field of haloes and their gas distribution by using
\citet{nfw} profiles (NFW)
\begin{equation}
\rho_{NFW} (R)= \frac{\delta \rho _c(z)}{\frac{R}{R_s} \left( 1+\frac{R}{R_s}
\right)^2},
\end{equation}
where the characteristic overdensity $\delta$ is given by:
\begin{equation}
\delta = \frac{200}{3} \frac{c^3}{F(c)}
\end{equation}
and where we choose the concentration parameter $c=10$ as our fiducial value.
The scale radius $R_s$ is the
ratio $R_s =R_{200}/c$ and the function $F(t)$ is given by:
\begin{equation}
F(t) = \log(1+t) - \frac{t}{1+t}.
\end{equation}

Inside haloes, the gas density is normalised to the total amount of hot gas
given by the semi--analytic recipes of \citet{semi}, that is $\rho_o =
\left( M_{hot} / M_{200}\right) \rho_{NFW}$.
For galaxies in groups or clusters, we follow the density profile of
the galactic halo until its density equals the density of the parent halo,
where the dynamics of the group becomes dominant.
Similarly, the density profile of the group is followed until the wind reaches
the point where the dark matter halo density becomes equal
to a fixed fraction of the mean universal density. After this point, the gas
density is assumed to be constant and equal to 0.8 times the baryonic mean
density.

The velocity of the surrounding medium $v_o$ is calculated assuming that the
gas dynamics is dominated by infall close to galaxies and by the Hubble flow at
larger distances. The outward $v_o$ is therefore given by the sum of two
contributions:
\begin{equation}
v_o(R) = - \frac{1}{2} v_{esc}(R) + H(z)\cdot R ,
\end{equation}
where the escape velocity $v_{esc}$ at a radius $R$ for a NFW profile is 
\begin{equation}\label{nostra}
v_{esc}^2(R) = \frac{2GM_{200}}{R} \frac{\log(1+ \frac{R}{R_s})}{F(c)}.
\end{equation}

When the velocity of the shock front equals the velocity of the intergalactic
gas, the dynamics of the wind joins the Hubble flow and no more mass is
accreted.

For winds expanding into the IGM, the pressure of the ambient medium $P_o$ is
given by the equation of state for an adiabatic gas, $P_o = c_s \rho_o
^{\gamma}$, where $c_s$ is the sound speed of the IGM and $\gamma = 5/3$ is the
adiabatic index for a monoatomic gas.
For winds expanding inside haloes, we make the assumption that the gas behaves
like an isothermal gas sphere with temperature \citep{nfw}
\begin{equation}\label{isotemp}
T_{o} = \frac{\mu m_H V_{200} ^2}{2k},
\end{equation}
where $\mu$ is the molecular weight, $m_H$ the hydrogen mass,
$V_{200}$ the virial velocity of the halo and $k$ the Boltzmann constant.
The pressure of the intracluster gas at radius $R$ can then be recovered from
the equation of state, which gives
\begin{equation}\label{isopressure}
P_o (R) = \frac{1}{2} \rho_o (R) V_{200} ^2 .
\end{equation}

The entrainment fraction $\varepsilon$ represents the fraction of the
medium surrounding the bubbles that is swept up by the wind and mixed into
the bubble fluid or accreted onto the shell. $\varepsilon$ is treated here as a constant free
parameter for simplicity, but in principle there is no reason why it should not
vary during the evolution of the winds.
What we are calling ``entrained'' gas is not the entrained ISM which
loads observed winds near their base, but is rather ambient gas which
has mixed into the hot bubble phase either through turbulent mixing of
shocked diffuse ambient gas or by evaporation and ablation of the
filament gas, most of which continues falling onto the galaxy. The
latter process is similar to that which is thought to load ISM mass
onto stellar winds and supernova blastwaves.
Both observations (e.g. \citealt{martin}) and simulations \citep{sh} suggest
that the entrained mass may be the dominant contribution to the hot gas mass
within a wind bubble.

The remaining fraction of mass $1 - \varepsilon$ is assumed to be in dense
clouds which are not entrained by the outflow.
A low value $\varepsilon < 1$ may reflect either a clumpy ambient
medium or a heavily fragmented wind, while $\varepsilon \sim 1$ describes
a near--homogeneous medium, which can be entirely swept up.
This parameter is of particular relevance because it plays a key
role in determining the fate of the wind: the mass accretion rate
depends on $\varepsilon$ and the larger the mass accreted onto the
wind, the bigger the energy required to accelerate it.
The ram pressure increases linearly with $\varepsilon$, again
influencing the energetics of the winds.
The net effect of an increase in the entrainment fraction is thus a decrease in
the shock velocity, which may lead to the collapse of the wind if
the energy input from the starburst is not sufficiently large.
Since the shock velocity is what ultimately determines how far into the
IGM the winds travel, a large variation of the volume filling factor is
expected as $\varepsilon$ varies. We will analyse this aspect further in section
\ref{quattro}.

\subsubsection{Wind Merging}
\label{shellmerging}

When two galaxies merge, we assume that also their winds ``merge''. If only
one galaxy is blowing a wind, then its wind will be attached to the merged
galaxy without modifications.
The merging of bubbles and shells is realised by assuming conservation of
volume, mass, momentum and energy.

Conservation of mass requires the final mass $m$ of the new bubble (shell) to be
the sum of the masses $m_1$ and $m_2$ of the two merging winds, that is $m = m_1
+ m_2$.
Similarly, the metal mass in the merged wind is $m_z = m_{1,z} + m_{2,z}$ and
the total energy $E = E_1 + E_2$.
Conservation of volume requires that the total volume $V$ of the final wind
cavity is equal to the sum of the volumes of the two single cavities $V = V_1 +
V_2$.
Since winds are spherical, the new shock radius is $R=\left( R_1 ^3 +
R_2 ^3\right)^{1/3}$.
The shock velocity is given by the conservation of momentum:
\begin{equation}
v_s=\frac{m_1 v_{s1} + m_2 v_{s2}}{m_1 + m_2}.
\end{equation}

\section{Evolution of Winds}
\label{tre}

\begin{figure*}
\includegraphics[width=\textwidth]{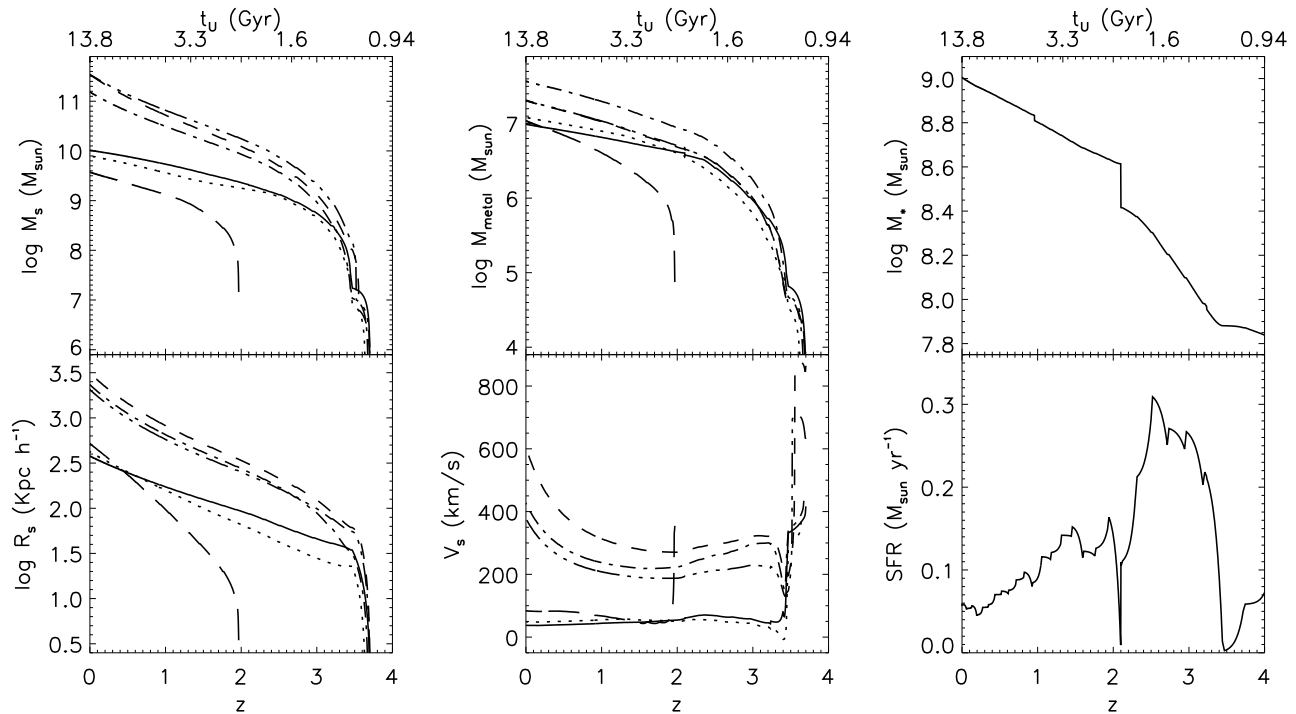}
\caption{Example of wind evolution for a dwarf field galaxy extracted from the
galaxy population of M3. From left to right and top to bottom, we show the
evolution of the bubble mass, the bubble metal mass, the stellar mass, the shock
radius, the shock velocity and the star formation rate of the galaxy.
The vertical jump in the stellar mass at $z\sim 2$ is due to a merging event,
which triggers the star formation activity of the galaxy.
$t_U$ is the age of the Universe in Gyr.
The lines represent different combinations of our model parameters (this
notation for the model parameters will be maintained throughout the paper):
(1) solid line: $K=0.5$, $\varepsilon = 1$;
(2) dotted line: $K=0.5$, $\varepsilon = 0.6$;
(3) dashed line: $K=0.1$, $\varepsilon = 0.3$;
(4) dashed dotted line: $K=0.5$, $\varepsilon = 0.3$;
(5) dashed three--dotted line: $K=1$, $\varepsilon = 0.3$;
(6) long dashed line: $K=0.5$, $\varepsilon = 0.1$.}
\label{two}
\end{figure*}

In this section we will discuss the evolution of winds and how our results
depend on our model parameters. In Subsection \ref{onegal} we show the
evolution of a wind emerging from a dwarf field galaxy and
in subsection \ref{allg} we focus on the population of wind bubbles and shells.
In Subsection \ref{trends} we discuss the general results and in Subsection
\ref{pm} we deal with bubbles and shells.

\subsection{Single Galaxy}
\label{onegal}

New--born winds expand initially inside the dark matter haloes in which their
host galaxies reside.
Since the amount of gas in haloes depends on the efficiency of cooling and
may vary significantly from galaxy to group, each wind has a ``personal''
history different from all the others. This history depends both on the
properties of the parent galaxy and on the environment where the wind expands.

In Fig. \ref{two} we show the evolution with time of a wind emerging from a
galaxy, chosen randomly from the galaxy population of M3, as a function of
our model parameters $K$ and $\varepsilon$.
The galaxy is a dwarf field galaxy first identified at $z\sim 10$, that remains
a ``central'' galaxy until $z=0$.
It forms stars in a rather continuous way throughout its lifetime, as
shown in the bottom right panel, but although the star formation rate never goes
to zero, it is generally too low to power a wind before $z\sim 4$.

This galaxy gives a few examples of some features that may appear during the
life of galaxies and winds.
For example, the vertical jump at $z\sim 2$ in the top right panel is due
to the merging of a satellite onto the galaxy, which contributes its stellar
mass to the central galaxy and somewhat triggers its declining star formation
activity.

Despite the rather weak star formation activity, the galaxy is able to power a
wind that finally escapes the gravitational attraction of the galaxy at $z<3$.
The galaxy attempts to blow winds since $z\sim 9$, but all the previous bubbles
(not shown in Fig. \ref{two}) are short lived and eventually collapse back onto
the galaxy because of their insufficient energy and momentum. This is
commonly known as a galactic fountain.
The formation of a wind powerful enough to escape the galactic potential
coincides with a burst of star formation at $z\sim 3.5$.
The evolution with time of the initially adiabatic bubble is strongly dependent
on the model parameters and on the different physics that they underline.
In some cases the wind remains adiabatic throughout its lifetime, while in
others it rapidly cools down and forms a shell of cold gas pushed by the
momentum of the ejecta.
Highly mass loaded winds need a larger energy input to power the expansion
and they are more likely to switch to the
momentum driven regime soon after blow out than less mass loaded bubbles.
These two cases are easily distinguishable in Fig. \ref{two}:
the momentum driven shells have the lowest shock velocites, the
smallest shock radii and the smallest wind masses, while the pressure driven
bubbles have the highest shock velocites, the largest radii and the highest wind
masses. At first order, the shock velocity is an indication of the temperature
of the bubble and it can be clearly seen how it determines the evolutionary
regime of the wind.
This result is consistent with the fact that momentum driven shells and pressure
driven bubbles follow different theoretical expansion laws, that is $R \propto
t^{1/2}$ for shells and $R \propto t^{3/5}$ for bubbles. In fact, we find
$v_{shell} < v_{bubble}$.

\subsection{Properties of the Wind Population}
\label{allg}

\begin{figure}
\includegraphics[width=8.4cm]{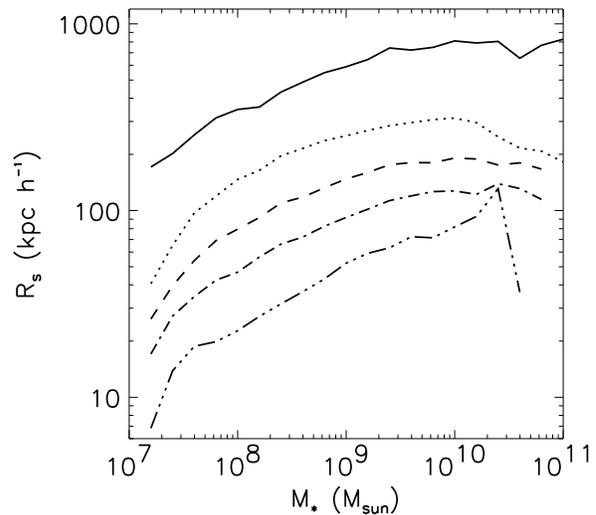}
\caption{The mean bubble radius for all the galaxies blowing a wind, as a
function of the stellar mass $M_{\star}$ of the host galaxy.
The data are shown for the model with $K=0.5$ and $\varepsilon = 0.3$
at different redshifts:
(1) solid line: $z=0$;
(2) dotted line: $z=1$;
(3) dashed line: $z=2$;
(4) dashed dotted line: $z=3$;
(5) dashed three--dotted line: $z=5$.}
\label{epsilon}
\end{figure}

Since the total number of galaxies in M3 is very large, we now calculate
``mean'' quantities to describe the global properties of the winds, and do not
focus further on individual cases.
We would like to point out that this necessarily gives a partial idea of the
whole picture, since there are no obvious correlations between the properties of
the galaxies and those of the winds.
However, we observe that the scatter in the distribution of the shock radii for
each model at a given redshift and as a function of the stellar mass is small
enough to assure that this quantity well represents the general trend for the
whole population.

\begin{figure}
\includegraphics[width=8.4cm]{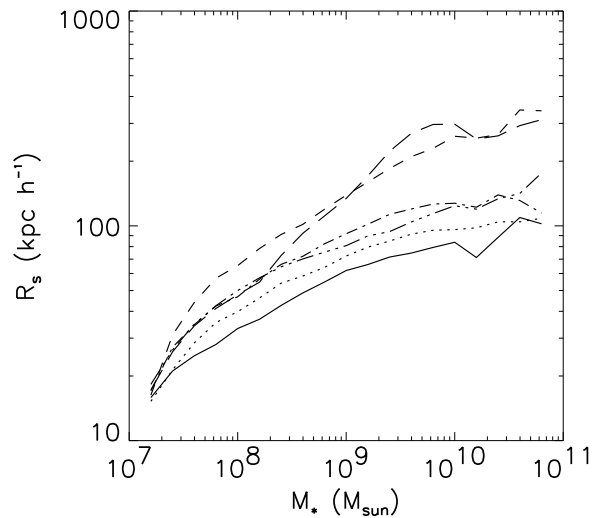}
\caption{The mean bubble radius for all the galaxies blowing a wind at $z=3$, as
a function of the stellar mass $M_{\star}$ of the host galaxy.
The lines correspond to different parameter choices as in Fig. \ref{two}.}
\label{kappa}
\end{figure}

Keeping all this in mind, in Fig. \ref{epsilon} we consider only wind--blowing
galaxies and we plot the mean values of the shock radii as a function of the
stellar mass $M_{\star}$ for different epochs of our model with $K=0.5$ and
$\varepsilon = 0.3$.
Clearly, the mean radius increases with time, as the winds have more time to
expand further from the galaxies.
For massive galaxies the mean radius appears to be considerably larger than
for less massive ones.
This effect has two explanations: first, these winds need higher
velocities to be able to escape from the gravitational attraction of their
haloes and therefore can cover larger distances; secondly, they often started
earlier in time.

In Fig. \ref{kappa} we plot the same quantity, but for different choices of
the model parameters at $z=3$. The scatter in the plot is large and the results
differ by as much as a factor of a few in the most extreme cases.
Our model parameters affect strongly the long term evolution of winds.
In particular, we find that more mass loaded bubbles tend to travel to shorter
distances than less mass loaded ones.
The distance to which a shock can travel depends crucially on the total amount
of mass accreted by the bubble and therefore on the fraction of the mass
entrained in the outflow as set by $\varepsilon$.

\begin{figure}
\hspace{-0.3cm}
\includegraphics[width=8.4cm]{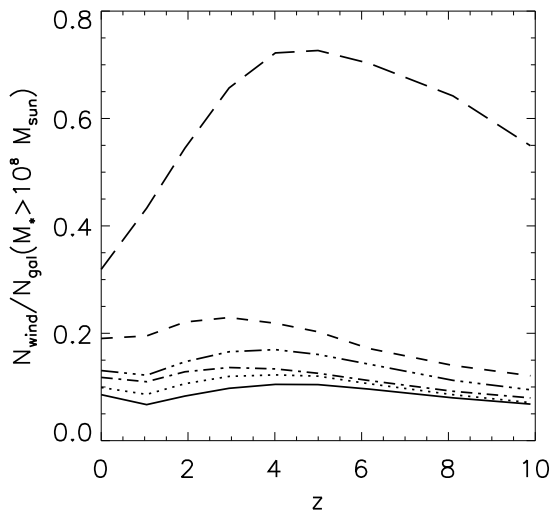}
\caption{The fraction of wind--blowing galaxies as a function of redshift and
model parameters. Here we consider only galaxies with $M_{\star} > 10^8$
M$_{\sun}$.
A large amount of swept up mass from the surrounding medium strongly
suppresses the ability of galaxies to power outflows.
The lines correspond to different parameter choices as in Fig. \ref{two}.}
\label{howmany}
\end{figure}

\begin{figure}
\includegraphics[width=8.4cm]{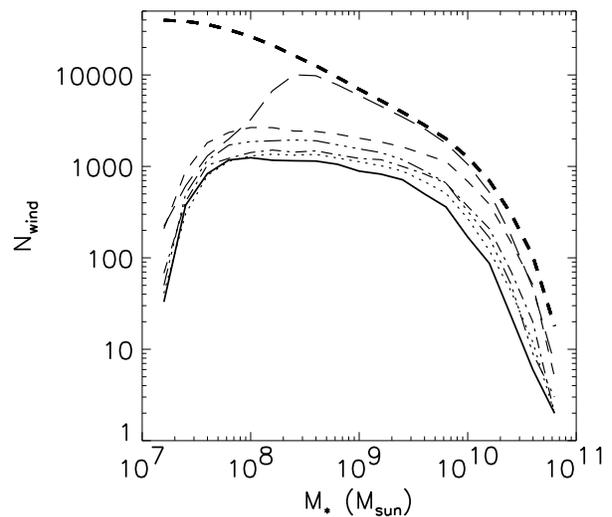}
\caption{The number of wind--blowing galaxies at $z=3$ as a function of the
stellar mass of the galaxy. For comparison, we have overplotted the total number
of galaxies as a function of stellar mass, as a thick dotted line.
The other lines correspond to different parameter choices as in Fig. \ref{two}.}
\label{ppp}
\end{figure}

In Fig. \ref{howmany} we show the fraction of galaxies with $M_{\star}> 10^8$
M$_{\sun}$ blowing a wind as a function of redshift and
in Fig. \ref{ppp} the number of galaxies with winds as a function of
stellar mass at $z=3$. Both quantities are plotted for different parameter
choices.
The number of galaxies blowing a wind depends strongly on the model parameters.
The entrainment fraction greatly affects the ability of galaxies to power a
wind and the overall effect of a high mass loading is to reduce the number of
wind--blowing galaxies by a significant factor. 
The suppression is particularly strong in galaxies with stellar masses
$M_{\star} \lesssim 10^8$ M$_{\sun}$, which dominate the stellar counts.

\subsection{General Trends}
\label{trends}

We find no clear connection between the present star formation
rate of a galaxy and the properties of the wind. The star formation
activity of a galaxy may switch off or decrease to very low values, while the
wind still has sufficient energy or momentum to escape the gravitational pull.
It is common in our simulations to find winds expanding in
the IGM a considerable time after the star formation activity and the energy and
momentum input from the source galaxy have ceased.

In principle, both bursts of star formation and quiescent star formation
may be able to power the winds, since we do not put any constraints on the star
formation rate to allow galaxies to blow winds.
It is not possible to predict \emph{a priori} when a wind will escape the
gravitational pull of a galaxy, since its evolution and its final fate are
linked to several factors, like the star formation and the mass accretion
history of the galaxy, the potential well of the dark matter halo in which it
expands, the amount of mass accreted both from the wind and the IGM and so on.
A bubble that is collapsing onto a galaxy may receive new
energy from increased star formation activity, triggered by mergers or gas
accretion, and it may start expanding again. On the other hand, an
expanding wind may start to collapse because its host galaxy falls into a
larger group and the gravitational attraction or the ambient pressure increases
by a large factor.
In models with a high entrainment fraction the energy input necessary to blow a
wind out of a galaxy is often too large to be provided by quiescent
star formation alone. On the other hand, quiescent star formation may succeed
to power outflows in galaxies residing in haloes for which the energy
required to overcome the gravitational attraction and the pressure forces of
the ambient medium is small.

Why can winds with low mass loading efficiency escape galaxies more efficiently
than more mass loaded ones? Why is this effect particularly strong in galaxies
with stellar masses in the range $10^8 < M_{\star} < 10^9$ M$_{\sun}$?
Let us consider equations (\ref{energy}) for the conservation of energy and
(\ref{label}) for the conservation of momentum.
A wind receives energy from the starburst and is slowed down by the
gravitational attraction of the central galaxy and by the ram pressure of the
ambient medium. Thermal pressure effects are consistent inside cluster haloes,
but are generally negligible in the IGM.
If the entrained mass is small, as in the case of $\varepsilon =0.1$, then a
large fraction of the bubble or shell mass is composed by the supernova ejecta
and the shocked ISM in the wind fluid, which are outflowing from the galaxy with
a velocity often much larger than the escape velocity of the galaxy.
Since little energy or momentum has to be spent by the wind to accelerate the
entrained mass, the shock velocity is less sensitive to energy losses by
pressure and gravity.
Such a wind has thus a higher probability to overcome the gravitational pull and
break free from the halo than more mass loaded winds.
When the mass loading is substantial, a significant part of the wind energy is
consumed to accelerate the entrained gas and the expansion slows down.
If the amount of energy spent to accelerate the swept up mass is
large compared to the total energy in the wind, the shock velocity may become
lower than the escape velocity of the galaxy. In this case, the wind cannot
escape and collapses back onto the galaxy.

In models with efficient mass loading, the suppression of winds is particularly
strong in galaxies with low stellar masses, because the delicate
momentum balance at blow out is easily dominated by losses by
pressure effects, which sum up to the ones by gravity.
To make the situation worse, the energy input in low mass galaxies is often not
as large as in more massive ones, due to a less intense star formation activity.
This may be why the formation of winds is suppressed in galaxies with $M_{\star}
\lesssim 10^8$ M$_{\sun}$ even in our model with $K=0.5$ and $\varepsilon =
0.1$.

From the initial conditions set by equations (\ref{cond1}) and (\ref{cond2}) we
see that the wind mass loss rate is proportional to the parameter $K$, which
implies that $K$ contributes to the mass loading of winds.
However, its overall contribution is smaller than the one set by the entrainment
fraction parameter, because $K$ only weakly affects the long term
evolution of the winds.

Mergers can provide a further key to understand why galaxies with intermediate
and large stellar masses do blow winds more efficiently than less massive ones.
Satellites falling onto central galaxies may be powering a wind whose bubble or
shell is accreted by the central galaxy.
These merged winds may receive a strong kick from the burst of star formation
that follows the merger and the resulting wind energy may be
high enough to allow the wind to escape the gravitational pull of the central
galaxy.
It is likely that if a massive galaxy is blowing a wind, then that wind started
before most of the halo mass was accreted and the wind was expelled to a large
radius at early times.
On the other hand, winds from massive galaxies may reach large distances in
relatively short times if they are powered by an intense burst of star
formation.

\subsection{Pressure Driven Bubbles and Momentum Driven Shells}
\label{pm}

We now want to investigate how pressure driven bubbles and momentum driven
shells coexist in our simulations and under which conditions a bubble
evolves into a shell.
Bubbles represent the first phase of the wind evolution, when the energy
provided by star formation makes a superbubble expand out of the disk of a
galaxy and into the galactic halo. This Sedov--Taylor phase is driven by the
pressure of the ejected hot gas, which accelerates and shock heats the ambient
medium crossed by the bubble.
The second phase of the wind evolution, that is the momentum driven outflow
phase, sets in when the cooling time of the shocked material becomes shorter
than the dynamical time of the wind and the material in the outer layers of the
bubble cools down and starts to accumulate in a thin shell.
During this phase, the work done to accelerate the entrained gas is radiated
away.

The cooling time of a bubble depends on several factors and in particular on its
density, temperature and metallicity. The lower the density and the higher the
temperature, the longer the cooling time, and viceversa. In practice, this means
that mass loaded bubbles evolving in a dense environment become radiative and
cool down after a very short timescale, while winds with inefficient mass
loading have a higher probability to remain pressure driven for a longer time.
Bubbles expanding out of dwarf field galaxies with shallow potentials tends to
remain adiabatic, while winds in large haloes quickly lose their energy to
counteract the pressure of the dense intracluster hot gas and are therefore
likely to become momentum driven soon after blowout, if they survive at all.

\begin{figure}
\hspace{-0.2cm}
\includegraphics[width=8.4cm]{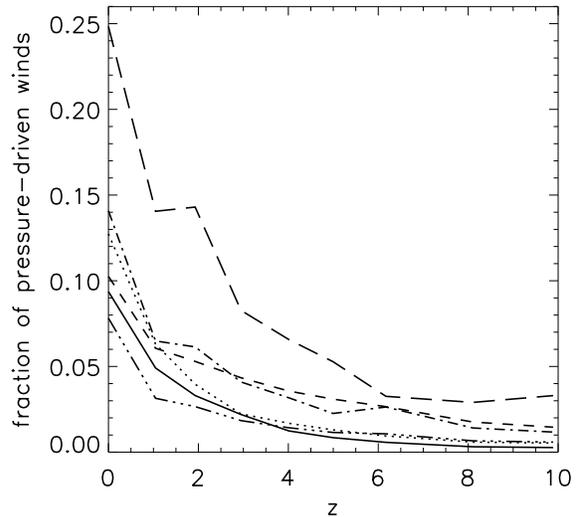}
\caption{The fraction of pressure driven bubbles as a function of redshift and
model parameters. Here we consider all the winds blowing from any kind of
galaxies and do not put any constraint on the minimum radius of bubbles and
shells.
The lines correspond to different parameter choices as in Fig. \ref{two}.}
\label{pdriven}
\end{figure}

In Fig. \ref{pdriven} we show the fraction of pressure driven winds as a
function of redshift for our different models.
It is evident at a first glance that at high redshift winds tend to be mostly
momentum driven, while at lower redshifts bubbles have a much higher probability
to remain adiabatic.
This is partly due to the higher mean density of the Universe during its infancy
and partly to a lower energy input from star formation, which determines lower
bubble temperatures and shorter cooling times immediately after blowout.
The cooling time of bubbles is therefore comparable to the age of the winds and
a cool shell often forms even before a wind has reached the virial radius of the
galactic halo.

In this graph we consider all the galaxies presently blowing a wind and we do
not put any constraint on the minimum radius of winds.
This is because we want to highlight how the transition from pressure driven
bubbles to momentum driven shells may be the first indication that a wind is not
powerful enough to escape the galaxy attraction.
In fact, when we plot only those winds with radii exceeding the virial radius of
the galaxy, that is $R > R_{200}$, we find that the fraction of pressure driven
bubbles increases by a factor of a few, with respect to the fraction of momentum
driven shells.
This indicates that pressure driven winds are overall more likely to escape
galaxies than momentum driven ones.

\begin{figure*}
\includegraphics[width=\textwidth]{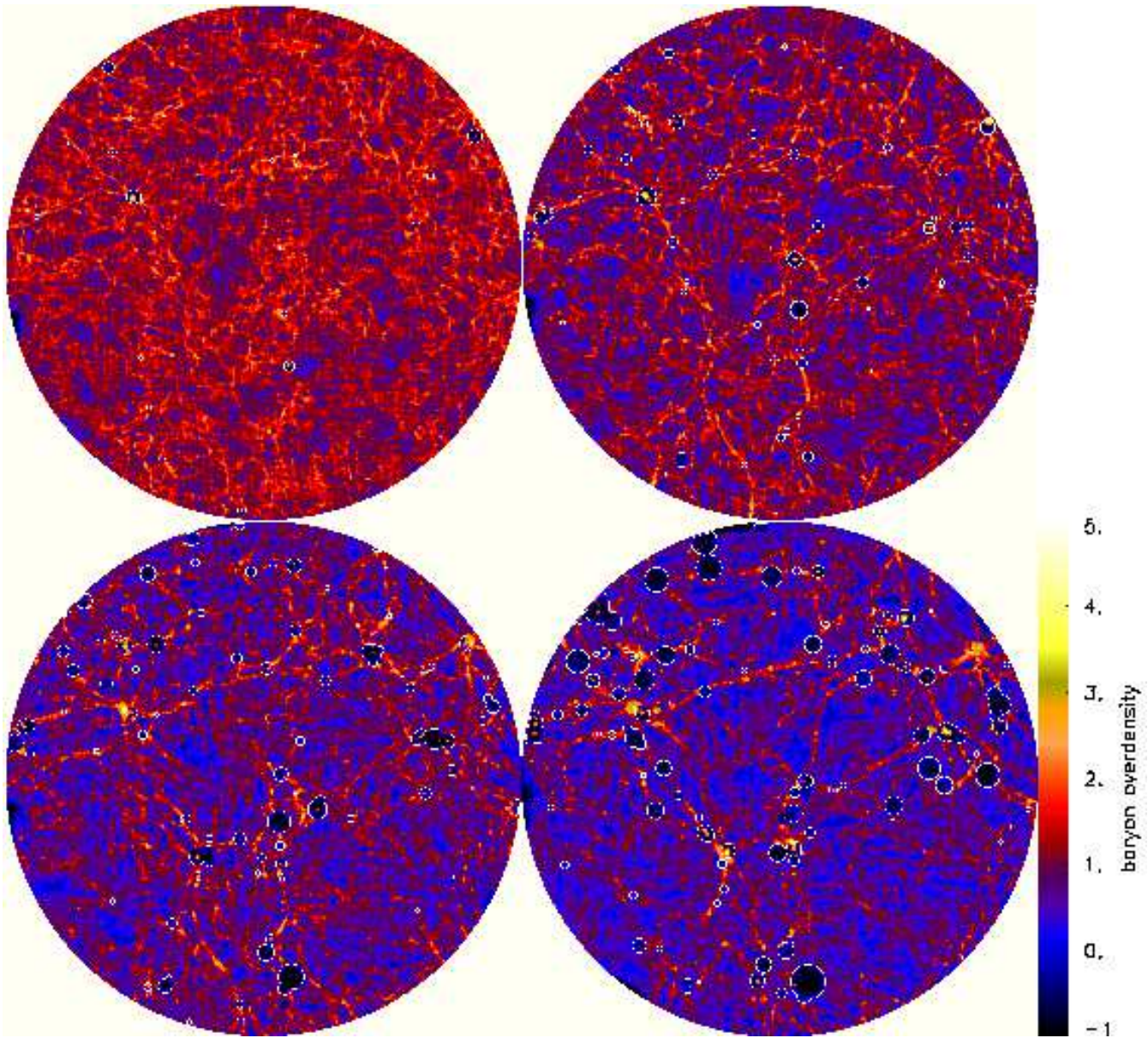}
\caption{Structure of wind--filled regions as a function of time.
The simulations have been realised for M3 assuming $\varepsilon =0.3$
and $K=0.5$.
From top to bottom and left to right, the redshifts of the presented
snapshots are respectively $z=5$, $z=3$, $z=2$ and $z=1$. The colour coding is
the same for all snapshots. The diameter of the region shown is $52 h^{-1}$
Mpc in comoving units. The white contours show the extension of winds at the
snapshot time, while the dark regions inside the countours are the regions
depleted of gas by outflows.}
\label{picture}
\end{figure*}

\section{Volume Filling Factor of Winds}
\label{quattro}

An estimate of the volume filling factor $f_v$ of galactic winds at the
redshifts where absorption in quasar spectra is observed can be translated into
an estimate of the probability to find disturbances in the Ly$\alpha$ forest due
to feedback effects and, in particular, to the presence of wind bubbles.
Disturbances here mean regions of the spectra where there are significant
variations in the optical depth, due to nongravitational processes stirring the
IGM.

Observationally, it is quite challenging to estimate $f_v$ with any
accuracy. Published estimates range from 0.003 to 40\%
(\citealt{heckman}, \citealt{cecil}, \citealt{rauch}) at $z\sim 3$. Surely the
large scatter is due to the fact that the estimates are mostly indirect and are
based on different, perhaps incompatible, assumptions, for example about the
geometry of the disturbances.

In Fig. \ref{picture} we show an example of the evolution of winds from $z=5$
to $z=1$. A thin slice is cut through the central plane of the simulation and
the density distribution of the gas in the slice is shown. The contours indicate
the surface of the winds and the black regions inside these contours
represent the regions which have been depleted of low density gas by winds.
To obtain these simulated distributions, we proceeded as follows.
First we recover the density of the gas by applying an
SPH smoothing to the distribution of the dark matter particles and by assuming
that the distribution of the gas follows the distribution of the dark matter.
We then identify the portion of space which is inside one or
more bubbles and we consider only those particles that are found
inside this region. Particles outside this region are of course not affected by
winds.
We compute the fraction $x$ of baryons removed from the IGM as the ratio of the
total baryon mass of the bubbles plus the galaxies to the total baryonic mass
initially associated with the dark matter inside bubbles.
We then tag the lowest density particles in the region
affected by winds until the same fraction $x$ of the enclosed mass is marked.
Finally, we cut a thin slice through the density distribution of the
remaining particles in our simulated region, as represented in Fig.
\ref{picture}.

\begin{figure}
\includegraphics[width=8.4cm]{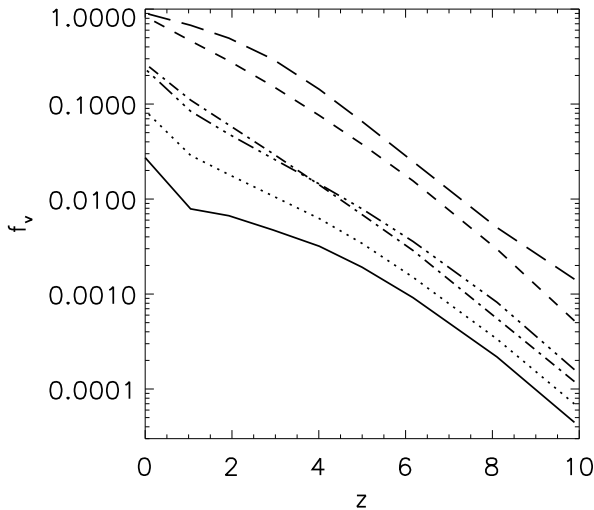}
\caption{Volume filling factor of winds in M3 as a function of redshift and
model parameters.
The lines correspond to different parameter choices as in Fig. \ref{two}.}
\label{timek}
\end{figure}

We calculate the filling factor of winds by superimposing a 3--dimensional grid
on our high resolution region and identifying all the grid points inside winds.
Since
$f_v$ represents the fraction of space occupied by winds, its value is given
simply by the ratio between the number of points flagged and the total number of
points in the grid. Note that this estimate ignores the mass fraction
$1-\varepsilon$ of the IGM which is in ``dense clouds'' and so avoided
entrainment.
We use a $512\times 512\times 512$ cubic grid, centered on the centre
of mass of the high resolution region and with a side of 52 $h^{-1}$ Mpc, but
we limit our analysis to a sphere of diameter 52 $h^{-1}$ Mpc. In principle,
a larger region with irregular contours could be identified.

The trends highlighted in paragraph \ref{trends} are recovered for the behaviour
of the volume filling factor, shown in Fig. \ref{timek} as a function of time
and parameters.
The smaller values for $f_v$ are clearly related to the cases where
fewer mass loaded bubbles are formed and expand into the IGM and viceversa.
By varying $K$ and $\varepsilon$, we can obtain a broad range of values of
$f_v$.

\begin{figure}
\hspace{-0.2cm}
\includegraphics[width=8.4cm]{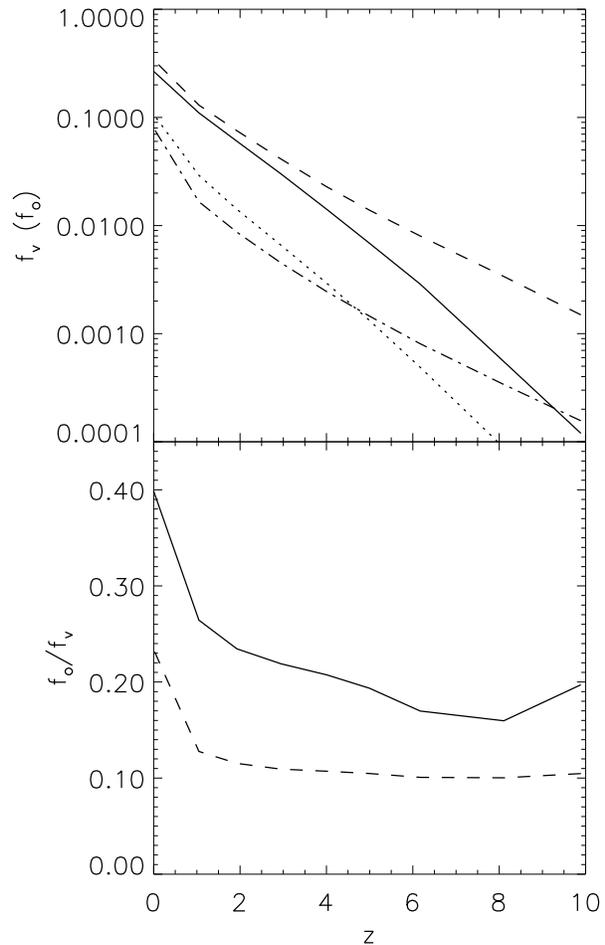}
\caption{
The top panel shows the volume filling factor $f_v$ and the overlapping $f_o$ of
winds for clustered sources and for randomly distributed sources.
The results are for our model with $K=0.5$ and $\varepsilon = 0.3$, that we
assume as our fiducial model here.
In the upper panel, the lines represent respectively:
(1) thick solid line: fiducial model;
(2) dotted line: overlapping for the fiducial model;
(3) dashed line: random positions;
(4) dashed dotted line: overlapping for the random positions model.
In the bottom panel we compare the ratio $f_o / f_v$ for clustered
sources (solid line) and for randomly distributed sources (dashed line).}
\label{over}
\end{figure}

In the model with $\varepsilon = 0.1$ and $K=0.5$, the fraction of volume
occupied by winds still increases steadily after $z\sim 2$, but, taking into
account the conversion between redshift and cosmic time, less strongly than
before.
This is probably due to the clustering of the wind sources, which becomes more
prominent at lower $z$.
As the galaxies cluster and the fraction of the volume occupied by winds and
shells increases, the probability of overlapping rises.
We do not model the overlapping of winds in a complete way, since our
one--dimensional approach does not allow us to take into account the
three--dimensional distribution of galaxies and winds on the sky, but we can
quantify the overlapping \emph{a posteriori}.
We define $f_o$ as the fraction of our
simulated volume which is reached by more than one wind. This definition is
analogous to the definition of $f_v$ and in practice $f_o$ can be evaluated
simply by counting the fraction of grid points that lie inside two or more
winds.

In Fig. \ref{over} we show the results of such a measurement for our model with
$\varepsilon = 0.3$ and $K=0.5$.
Because the galaxies are associated in groups, the wind cavities occupy a
smaller fraction of space than they would if they were randomly
distributed. At the same time, winds can
run into each other much more easily, so the overlapping becomes
significant already at high redshift, as is shown in the bottom panel of Fig.
\ref{over}. The ratio $f_o /f_v$ represents the fraction of the cavity volume
which is reached by two or more winds. While for randomly distributed galaxies
overlapping is more uncommon, for clustered
galaxies the volume with overlapping winds is already twice as big as in the
Poisson case at $z\sim 10$. The probability that winds overlap significantly
increases in models with high filling factors.

\section{The Wind Mass Budget}
\label{budget}

A second important indicator of the impact of winds on the surrounding
medium is the fraction of intergalactic gas that they affect.
This ``wind mass fraction'' ($f_m$, hereafter) is directly
dependent on the entrainment fraction and on the mass of gas ejected by the
galaxies. $f_m$ also depends in a crucial way on the density of the ambient
medium crossed by the wind, which, together with the entrainment fraction,
determines the accretion rate of gas of winds. In fact, winds expanding
into high density regions like filaments or groups may entrain, and therefore
affect, much more mass than winds from field galaxies, which are normally
embedded in a lower density environment.

\begin{figure}
\includegraphics[width=8.4cm]{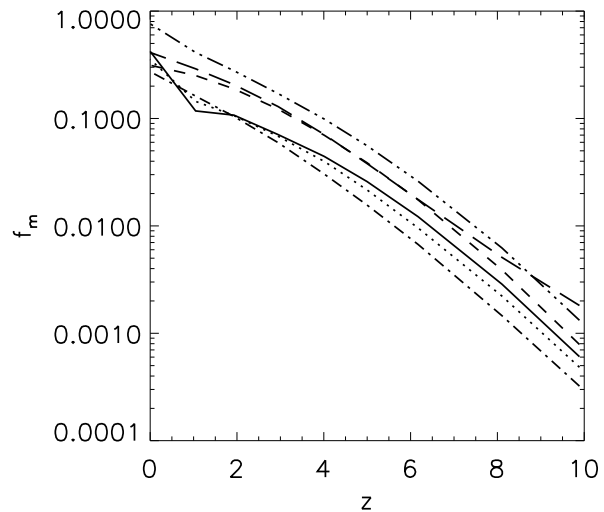}
\caption{The fraction of IGM mass affected by winds $f_m$ as a function of
redshift and model parameters.
The lines correspond to different parameter choices as in Fig. \ref{two}.}
\label{massfill}
\end{figure}

We estimate $f_m$ as the ratio of the mass in winds to the total mass of IGM
in our simulated box. Our results are presented in Fig. \ref{massfill}. 
While $f_v$ is determined only by the physical extension of winds, $f_m$ is
somewhat more difficult to estimate. In fact, one has to deal correctly with
overlapping, which is not treated self--consistently in our
semi--analytic prescriptions.
To do this, we have to correct approximately for the fact that in our spherical
wind model the same material can effectively be swept up two or more times
when winds overlap.
We first calculate the total mass of IGM inside wind cavities in two different
ways, that is from the dark matter particle distribution ($m_{igm, p}$) and from
our semi--analytic prescriptions for the distribution of gas around galaxies,
given in subparagraph \ref{environment}, ($m_{igm, sa}$).
The first method reflects the ``real'' 3--dimensional distribution of matter in
our simulated region. We then define $y$ as the ratio between the two,
that is $y = m_{igm, p} / m_{igm, sa}$.
The bubble mass, defined in equation (\ref{monster1}), is the sum of the mass
from supernova ejecta (plus the shocked ISM) and of the gas mass entrained along
the way $m_e$, that is: $m = m_w + m_e$.
The mass of supernova ejecta $m_w$ is independent of overlapping effects.
Conversely, the entrained mass $m_e$ does depend on overlapping and we thus
rescale it by the factor $y$ to obtain the actual swept up mass $m'_e = y m_e$.
The rescaled wind mass is thus $m' = m_w + m'_e$. Finally, we calculate the
fraction of IGM mass affected by winds as $f_m = m' / m_{igm, p}$.

The fraction of mass affected by winds $f_m$ varies differently from the volume
filling factor $f_v$ and, in particular, it depends less strongly on the values
of the model parameters.
The wind mass fraction is lower than 10\% at $z>4$, and at $z\sim 2$
is still not higher than 30\%. This suggests that winds are unlikely to
significantly modify the properties of the \lya\ forest at $z \sim 3$ and it is
a clear indication that the \lya\ forest itself is not entirely
modelled by the mechanical effects of feedback.
By comparing Fig. \ref{massfill} and \ref{timek}, it appears
that at low redshifts the winds with low mass loading efficiency
have in some cases a large $f_v$ and a small $f_m$, or viceversa.
$f_m$ is normally higher than the volume filling factor at any redshift,
indicating that although winds do not travel far into the IGM, the effective
amount of mass affected may be large.
This effect is easy to understand, when one considers that most of the mass in
the Universe lies in proximity of high density regions like filaments and
clusters and that most of the winds are located in these same regions of space.
Only in a few models the volume filling factor reaches higher values than the
fraction of mass in winds. In these cases, the actual fraction of intergalactic
mass affected by outflows is small even when the winds physically fill a
large region of space. This effect is associated with the models with the
lowest mass loading efficiencies.

\section{The Ejection of Metals}
\label{ejection}

\begin{figure}
\hspace{-0.2cm}
\includegraphics[width=8.4cm]{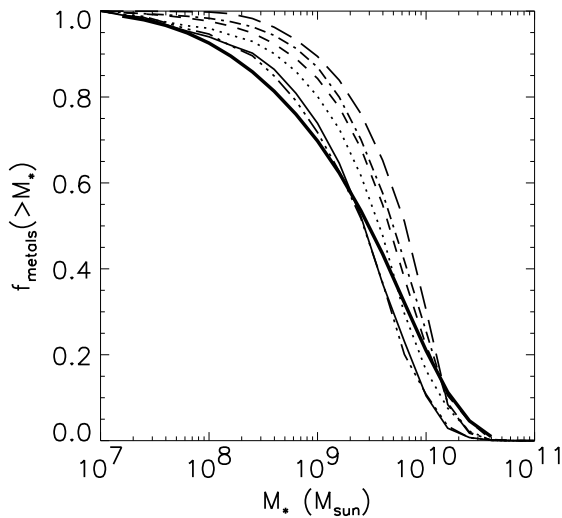}
\caption{The cumulative distribution of the metal mass in winds at $z=3$
as a function of the stellar mass of the parent galaxies.
For comparison, the cumulative distribution of the stellar mass in galaxies is
overplotted as a thick line.
The other lines correspond to different parameter choices as in Fig. \ref{two}.}
\label{zmet}
\end{figure}

Which galaxies eject the metals we observe in the IGM?
To answer this question, in Fig. \ref{zmet} we plot the cumulative
distribution of metal mass in winds as a function of the stellar mass of the
ejecting galaxies for our different models at $z=3$.
For comparison, we show the cumulative distribution of the stellar mass in
galaxies as a thick straight line.

Different combinations of our model parameters lead to somewhat different shapes
for the distribution, but the main conclusion we can draw from Fig. \ref{zmet}
is that at $z=3$ most of the metals are ejected by galaxies with stellar masses
in the range $10^8 - 10^{10}$ M$_{\sun}$, which roughly corresponds to the
mass range of dwarf galaxies.
Galaxies with stellar masses larger the $10^{10}$ M$_{\sun}$ or smaller than
$10^8$ M$_{\sun}$ do not significantly contribute to the pollution of the IGM.
In fact, these galaxies eject altogether about only 20\% or less of the metals
in winds whose radii exceed the virial radius of the source galaxy.
About 80\% of the metals come from galaxies with stellar masses in the
range $10^8 < M_{\star} < 10^{10}$ M$_{\sun}$. In all models, galaxies with
$10^9 < M_{\star} < 10^{10}$ M$_{\sun}$ eject about 60\% of the metals, and
are therefore the main contributors to the pollution.
For comparison, about 80\% of the total stellar mass in our simulated region
lies in galaxies with $10^8 < M_{\star} < 2\cdot 10^{10}$ M$_{\sun}$.
Galaxies with stellar masses larger the $10^{10}$ M$_{\sun}$ contain about 20\%
of all the stars, but only in models with inefficient mass loading they
contribute up to 30\% of the ejected metals.

\begin{figure}
\hspace{-0.2cm}
\includegraphics[width=8.4cm]{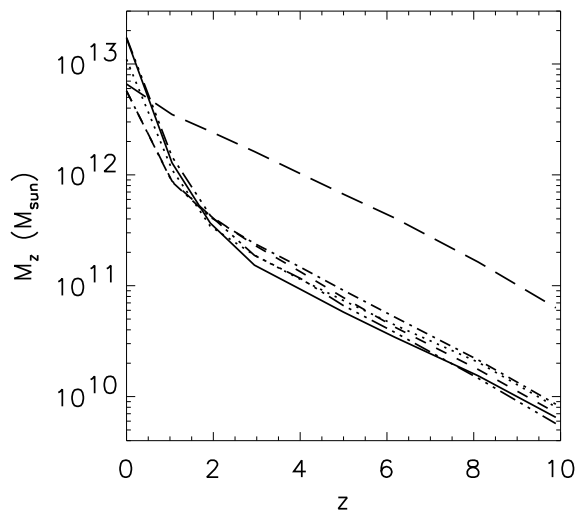}
\caption{The total mass of metals ejected by winds as a function of redshift and
model parameters.
The lines correspond to the different parameter choices as in Fig. \ref{two}.}
\label{lic1}
\end{figure}

In Fig. \ref{lic1} we show the total mass of metals ejected by winds as a
function of redshift. This mass increases steadily with decreasing redshift and
indicates that galaxies actively contribute to the metal enrichment of the IGM
throughout the history of the Universe.
The model with the lowest mass loading efficiency ($\varepsilon = 0.1$, long
dashed line) favours an intense ejection already at high redshift, while all the
other models tend to suppress it until more recent times. This behaviour is a
consequence of the suppression of winds in models with higher mass loading
efficiencies that we discussed in Figs. \ref{howmany} and \ref{ppp}.

\begin{figure}
\hspace{-0.5cm}
\includegraphics[width=8.4cm]{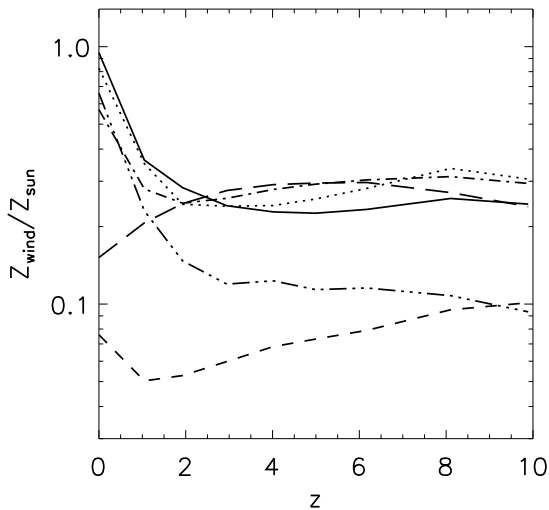}
\caption{The mean metallicity of winds as a function of redshift and model
parameters.
The lines correspond to different parameter choices as in Fig. \ref{two}.}
\label{lic2}
\end{figure}

In Fig. \ref{lic2} we show the mean metallicity of winds in solar units as a
function of redshift.
This depends on the ratio between the mass entrained from the
metal--free IGM and the metal rich gas accreted directly from the wind.
We do not find that the metallicity of winds increases with decreasing
mass loading efficiency, as it would be reasonable to expect, because the winds
forming in models with high mass loading efficiencies are also the ones which
travel the shortest distances in our simulations (cfr. Fig. \ref{kappa}),
and can therefore entrain the smallest amounts of metal--free ambient medium
to dilute their metal content and lower their metallicity.
Instead, we observe that the wind metallicity does depend on the parameter
$K$, which determines the quantity of ISM blown out of galaxies together with
the SN ejecta. This sets directly the metallicity of the wind fluid and,
consistently, we find that a low value of $K$ corresponds to metal--poor winds,
while a high $K$ to metal--rich ones.

\subsection{Metals in the Intergalactic Medium}

At $z\sim 2$ to 3, recent estimates of the IGM metallicity in regions with
densities close to the mean baryon density give values in the range $Z_{IGM}
\sim 10^{-3.5} - 10^{-2.5} Z_{\odot}$ (\citealt{simcoe}, \citealt{joop2}).
We can use the metallicity of our winds to attempt a very rough estimate
of the metallicity of the IGM in our simulated region, by multiplying
the fraction of intergalactic gas affected by winds by the wind metallicity.
As a result, we find that in our simulations the IGM
reaches metallicities in the range $Z_{IGM} \sim 10^{-2} - 10^{-1.2} Z_{\odot}$
between $z=3$ and $z=2$. The actual value of the metallicity depends on the
model parameters and may vary within a factor of a few.
The models with $\varepsilon > 0.1$ produce the lowest metallicities, while
the model with the lowest mass loading efficiency ($\varepsilon = 0.1$) predicts
the highest metallicity. This can be easily understood, since as we show in Fig.
\ref{lic1} this model predicts the highest amount of metals ejected into the
IGM at $z>1$.

Our estimates of the IGM metallicity are in excess of the observed values
by a factor of 10--100, depending on the model. A similar conclusion has
recently been formulated by \citet{cen}.
There are some possible explanations for this:
i) our simulated winds may expel more metals into the IGM  than actually happens
in real galaxies;
ii) the metals in winds do not effectively pollute the IGM uniformly, so that
different concentrations of metals could be found at different locations in
space;
and finally, iii) the ejected metals may not be efficiently mixed into the
observed ``cold'' gas, and therefore may not be detectable in absorption in the
\lya\ forest.

While our models might overestimate the total amount of mass and metals ejected
by galaxies by a factor of up to a few, it is unlikely that such a correction
would change our conclusion that the mean IGM metallicity is higher
than observations of the \lya\ forest suggest.

Of course, regions affected by winds may have metallicities of order $Z_{wind}$,
while unaffected regions would likely mantain their pristine chemical
composition.
Other mechanisms different from galactic winds may in principle pre--enrich
the lowest density regions of the Universe at earlier times, like e.g.
Pop III stars or pregalactic outflows. 
On the other hand, the high metallicities predicted by our simulations in the
outskirts of galaxies may explain why systematic velocity shifts of few
hundred km s$^{-1}$ between \lya\ and metal absorption lines have been found in
several quasar spectra.

Alternatively, the discrepancy between our estimate of the IGM metallicity and
the observed values cited above may be explained by the non--detection of part
of the intergalactic metals.
But why should we be unable to detect these metals?
The C IV and O VI detected in the spectra of quasars reside in a photoionised
gas at temperatures of about $10^4 - 10^5$ K. For higher temperatures,
collisional ionisation becomes efficient, so that carbon and oxygen are fully
ionised and do not absorb the UV photons anymore.
The temperature of winds may thus be a key factor to determine the observability
of their metal content in absorption.

\begin{figure}
\hspace{-0.5cm}
\includegraphics[width=8.4cm]{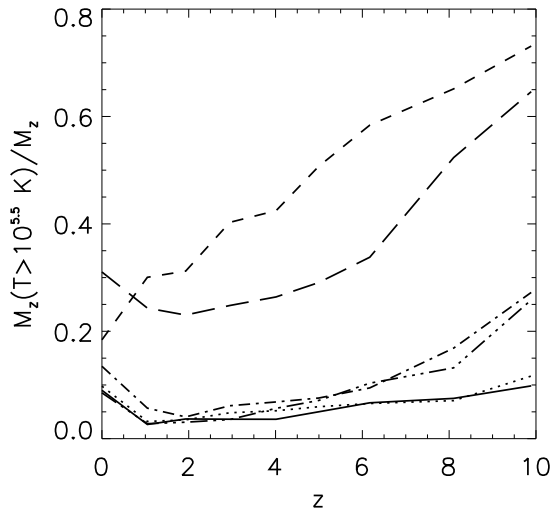}
\caption{The fraction of metal mass in winds with $T > 10^{5.5}$ K as a function
of redshift and model parameters. 
The lines correspond to different parameter choices as in Fig. \ref{two}.}
\label{hotfraction}
\end{figure}

In our model, we find that at $z=3$ a large fraction of the winds that have
escaped the potential wells of haloes have temperatures
lower than about $10^6$ K.
If the wind temperature drops below about $10^{5.5}$ K,
photoionisation replaces collisional ionisation as the main ionisation
mechanism and cooled shells could produce C IV and O VI absorption.
In most cases, pressure driven bubbles do have temperatures higher than $10^6$ K
and no metal absorption could take place.
In Fig. \ref{hotfraction} we plot the fraction of the metal mass transported by
winds with temperatures higher than $10^{5.5}$ K, which therefore would not
produce any observable absorption in the spectra of quasars.
Although this result is strongly
parameter dependent, it is clear that at $1\lesssim z\lesssim 5$ a significant
fraction of the metals resides in a hot gas that would leave no footprint
in the \lya\ forest.
This result may support the idea that the IGM is enriched to a higher level than
\lya\ observations can prove, but that the metals blown out of galaxies by
galactic winds are in many cases too hot to produce any detectable absorption in
the spectra of quasars.

Gas at temperatures as high as $10^7$ K is expected to emit radiation in the
X--ray band and one would expect to find X--ray emission in the IGM not
associated with jets or collapsed objects.
Indeed, X--ray emission from highly ionised metal species (O VIII and Ne X) in a
warm--hot IGM (WHIGM) may have been recently discovered by {\small CHANDRA}
(e.g. \citealt{nicastro}, \citealt{kernan}).
This hot gas may be shock--heated by galactic winds as well as from the process
of structure formation or jets from active galaxies.
If the first case is true, this gas may represent the hot metal enriched gas
in our bubbles, which is too highly ionised to produce absorption in
the Ly$\alpha$ forest.

\section{The effect of mass resolution}
\label{mares}

We want now to investigate the effect of the resolution in mass of our
N--body simulations in determining the volume filling factor and the fraction of
IGM mass affected by winds. 
To do this, we compare the results obtained from our four sets of simulations
with increasing mass resolution.
While a large population of dwarf galaxies is already forming at $z\lesssim 20$
in M3, only a few objects are assembling in M2 at the same epoch and in the
lower resolution runs M1 and M0 the first galaxies appear only at $z<15$ and
$z<7$, respectively.
The total number of galaxies in M3 is five times as large as in M2 at $z=0$ and
the number of galaxies with winds two times as large.

\begin{figure}
\includegraphics[width=8.4 cm]{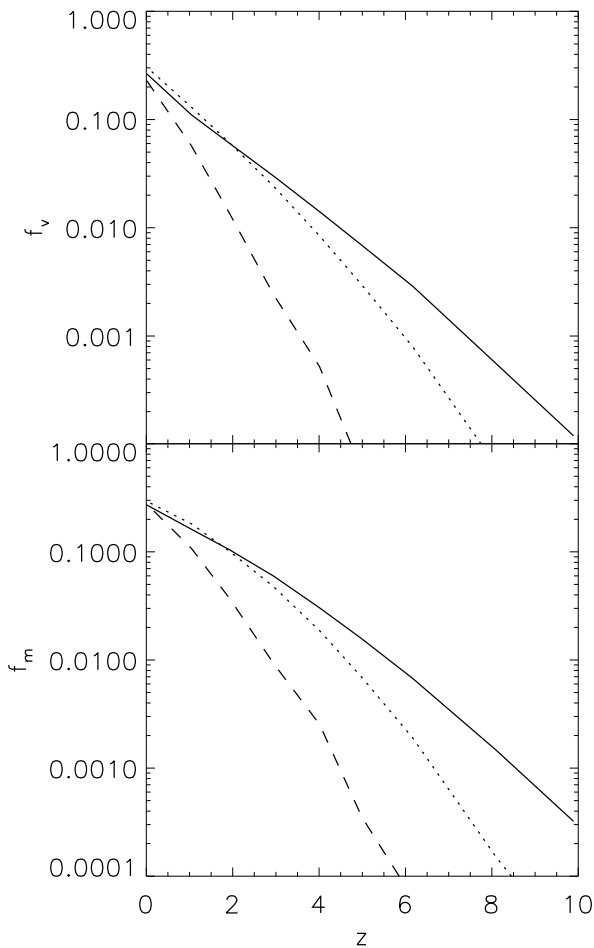}
\caption{
Dependence of the volume filling factor $f_v$ (upper panel) and of the fraction
of mass affected by winds $f_m$ (lower panel) on the mass resolution of the dark
matter N--body simulations.
The results are for a model with parameters $\varepsilon =0.3$ and $K=0.5$ and
the lines represent respectively three of our simulation sets:
(1) solid line:  M3, $M_p = 1.7\cdot 10^8 h^{-1}$ M$_{\sun}$;
(2) dotted line: M2, $M_p = 9.5\cdot 10^8 h^{-1}$ M$_{\sun}$;
(3) dashed line: M1, $M_p = 4.8\cdot 10^9 h^{-1}$ M$_{\sun}$.
The large galaxies resolved in M0 do not give any significant contribution  at
any redshift.}
\label{mfilling}
\end{figure}

In Fig. \ref{mfilling} we show the volume filling factor (top panel) and the
fraction of mass affected by winds (lower panel) in the M1, M2 and M3 simulation
sets. The results are shown for a model with parameters $\varepsilon =0.3$ and
$K=0.5$.
We find that at high redshift $f_v$ and $f_m$ strongly depend on the mass
resolution and on the ability to resolve galaxies which form in haloes with
total masses of about $10^9 - 10^{10}$ M$_{\sun}$.
These galaxies, only resolved in M3, include galaxies with both low and
intermediate stellar masses, field galaxies and satellites.

At $z\sim 0$ the results for the different sets of simulations converge.
This is an indication of the fact that dwarf galaxies are mainly responsible
for the pollution of the IGM at high redshift, but that their relative
contribution becomes smaller at very low redshift, when more massive galaxies
become the main sources of powerful winds and can account for most of the mass
and metals ejected into the IGM.

Despite the convergence of the global star formation in our simulated region
(cfr. Subsection \ref{sfhistory}), the star formation history of single objects
may vary from M2 to M3.
This determines a different evolution of winds with time: in M2 bubbles
tend to form at later times and expand faster than in M3, as a consequence of a
more intense star formation activity concentrated at later times and triggered
by a faster accretion of gas onto the fewer galaxies.
When bubbles form in M2, the dark matter haloes in which they expand already
possess total masses larger than the haloes in M3, but the larger energy
provided by star formation often compensates for the increased gravitational
attraction.
This is why the convergence of the curves in Fig. \ref{mfilling} is not exact,
but may differ by a small factor, generally not larger than a few percent.

One may ask if objects with stellar masses lower than about $10^7$--$10^8$
M$_{\sun}$ might give a substantial contribution to the pollution of the IGM
at redshifts where larger objects have not yet assembled.
Indeed, \citet{mfr} claim that the IGM has been polluted by outflows
from pregalactic objects, with total masses well below $10^8 - 10^9$ M$_{\sun}$.
In principle, winds may escape very easily the shallow potential wells of such
objects, if the energy input from star formation is high enough to accelerate
the accreted mass to velocities larger than the escape velocities of their
haloes.
At lower redshifts, the evolution of objects with total masses lower than $10^9$
M$_{\sun}$ may be affected by feedback effects that inhibite their star
formation activity (e.g. \citealt{hrl}, \citealt{mac}).
As a result, these objects would be unable to blow winds, making their
contribution to the pollution of the low redshift IGM negligible with respect to
other galaxy populations with higher stellar masses.
Unfortunately, our simulations do not have sufficient resolution to follow the
evolution of these objects. However, in the light of the results displayed in
Fig. \ref{mfilling}, we believe that such low stellar mass galaxies would not
significantly change our conclusions.

\section{Conclusions}
\label{cinque}

We have presented semi--analytic simulations of galaxy formation in a
cosmological context, which include the physics of galactic winds. The
semi-analytic prescriptions are applied to high resolution N--body simulations
of a typical ``field'' region of the Universe.

The results of our model can be quite accurately interpreted as a consequence
of the mass loading efficiency of winds.
The mass accumulated in bubbles is directly linked to the amount of
mass entrained from the ambient medium, set by the parameter $\varepsilon$, and
the ultimate fate of winds is strongly dependent on this swept--up mass.
Bubbles that load little mass from the surrounding medium can escape the
gravitational potential well of their host haloes more efficiently at every
redshift.
These bubbles are mostly composed of metal rich supernova ejecta and
shocked ISM and need to spend little of their energy to accelerate the
accreted gas. Since most of the energy injected by the starburst is available
to power the expansion of the bubble, these winds have the highest probability
to escape the gravitational attraction of haloes and expand into the IGM.
The formation of highly mass loaded winds is instead suppressed in all kinds of
galaxies, although the suppression is particularly strong in galaxies with
$M_{\star} \lesssim 10^{9}$ M$_{\sun}$. This is because the energy provided by
star formation is not sufficient to overcome the ram pressure of the infalling
material which adds to the gravitational pull of the galaxy.

Our estimates of the volume filling factor of winds (Section \ref{quattro}) and
of the fraction of IGM mass affected by winds (Section \ref{budget}) suggest
that galactic outflows are unlikely to significantly modify the properties of
the Ly$\alpha$ forest.
No obvious correlation is found between $f_m$ and $f_v$.
The volume filling factor is clearly dependent on the mass loading efficiency of
bubbles, with low values of $f_v$ associated with highly mass loaded bubbles and
viceversa.
The fraction of IGM mass affected by winds is usually comparable to the volume
filling factor.
Only in models with high mass loading efficiency we find that $f_m > f_v$, which
implies that the actual fraction of intergalactic mass affected by outflows may
be large even when the winds physically fill a small region of space.
This is a consequence of the clustering of matter on large scales and of the
fact that galaxies form in high density regions, where
their winds can sweep up a larger amount of material than they would if they
were expanding inside a low density region.

The efficiency of winds in seeding the IGM with metals is investigated in
section \ref{ejection}.
Galaxies with $M_{\star} \lesssim 10^{9}$ M$_{\sun}$
play a role in the chemical enrichment of the IGM only at very high redshifts,
when larger objects have not yet assembled.
At $z=3$ most of the metals are ejected by galaxies with $10^9 \lesssim
M_{\star} \lesssim 10^{10}$ M$_{\sun}$, while galaxies with $M_{\star} \gtrsim
10^{10}$ M$_{\sun}$ contribute only about 10\%--20\% of the ejected metals.
The result that metals are mostly ejected by relatively small galaxies
qualitatively agrees with the predictions of e.g. \citet{tom}, \citet{tv},
\citet{violence}.

Our estimates of the mean metallicity of the IGM are significantly higher than
the observed values at $z\sim 1$ to $z\sim 5$ and we have argued that metals in
the IGM might not be observable in absorption in the spectra of quasars because
of the high temperatures of winds.
In a forthcoming paper we will discuss the possibility of finding observable
signatures of cooled wind shells in the Ly$\alpha$ forest.

\section*{Acknowledgments}

We would like to thank V. Springel and the referee M.-M. MacLow for
useful discussions.
S.B. was partially supported by a Marie Curie fellowship by the European
Association for Research in Astronomy under contract HPRN--CT--2000--00132 and 
by a grant ``Progetto Giovani Ricercatori'' of the University of
Torino and is thankful to the Max Planck Institut f\"ur Astrophysik for the
kind hospitality and to C. Rickl, K. O'Shea, G. Kratschmann and M. Depner
for making life easier.
This work has been supported by the Research and Training Network ``The Physics
of the Intergalactic Medium'' set up by the European Community under contract
HPRN--CT--2000--00126.

\bsp

\label{lastpage}

\end{document}